\documentclass[]{spie}  

\usepackage{amsmath,amsfonts,amssymb}
\usepackage{graphicx}
\usepackage{multirow}
\usepackage[colorlinks=true, allcolors=blue]{hyperref}

\title{A Multi-rater Comparative Study of Automatic Target Localization Methods for Epilepsy Deep Brain Stimulation Procedures}

\author[a]{Han Liu}
\author[b]{Kathryn L. Holloway}
\author[c]{Dario J. Englot}
\author[d]{Benoit M. Dawant}
\affil[a]{Dept. of Computer Science, Vanderbilt University, Nashville, TN 37235, USA}
\affil[b]{Dept. of Neurosurgery, Virginia Commonwealth University, Richmond, VA 23284, USA}
\affil[c]{Dept. of Neurosurgery, Vanderbilt University Medical Center, Nashville, TN 37235, USA}
\affil[d]{Dept. of Electrical and Computer Engineering, Vanderbilt University, Nashville, TN 37235, USA}

\begin{document} 
\maketitle

\begin{abstract}

\end{abstract}
 Epilepsy is the fourth most common neurological disorder and affects people of all ages worldwide. Deep Brain Stimulation (DBS) has emerged as an alternative treatment option when anti-epileptic drugs or resective surgery cannot lead to satisfactory outcomes. To facilitate the planning of the procedure and for its standardization, it is desirable to develop an algorithm to automatically localize the DBS stimulation target, i.e., Anterior Nucleus of Thalamus (ANT), which is a challenging target to plan. In this work, we perform an extensive comparative study by benchmarking various localization methods for ANT-DBS. Specifically, the methods involved in this study include traditional registration method and deep-learning-based methods including heatmap matching and differentiable spatial to numerical transform (DSNT). Our experimental results show that the deep-learning (DL)-based localization methods that are trained with pseudo labels can achieve a performance that is comparable to the inter-rater and intra-rater variability and that they are orders of magnitude faster than traditional methods.

\keywords{Epilepsy, Deep Brain Stimulation, Anterior Nucleus of Thalamus, Localization, Heatmap Matching, DSNT}

\begin{figure}[b]
\includegraphics[width=1\columnwidth]{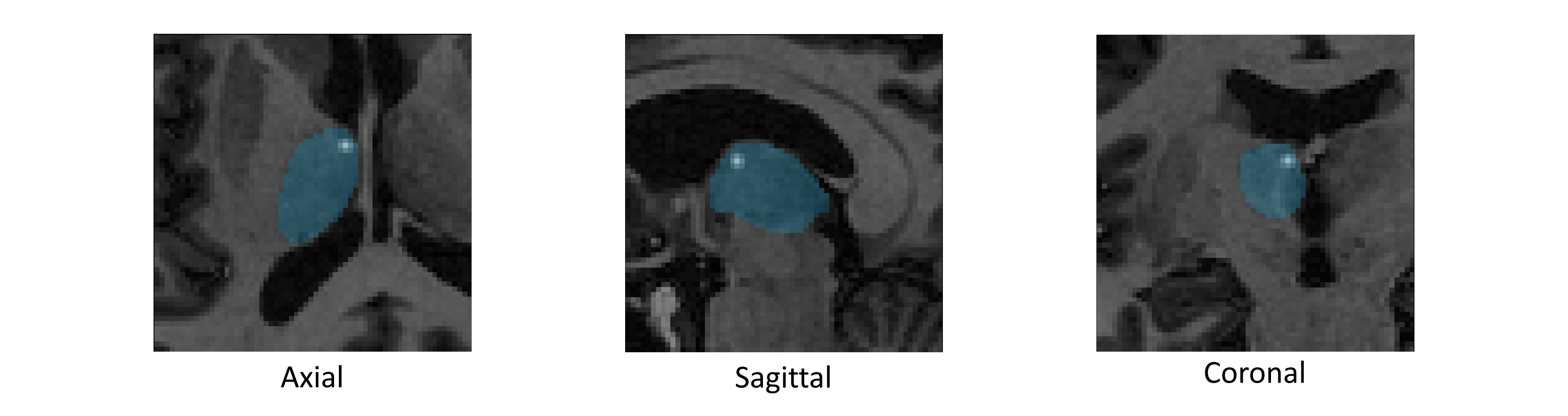}
\centering
\caption{An example of ANT-DBS target on the right thalamus. The thalamus mask is displayed in blue, and the target is denoted as the bright spot.}
\label{fig1}
\end{figure} 

\section{Introduction}
Epilepsy is one of the leading neurological causes of loss of quality-adjusted life years and affects 65 million people worldwide\cite{bg1}. It has been shown that despite the use of first-choice anti-epileptic drugs and satisfactory seizure outcome rates after resective epilepsy surgery, a considerable percentage of epilepsy patients suffer with persistent seizures\cite{bg2}. ANT-DBS has emerged as an alternative treatment option for these patients. DBS procedures involve the delivery of a predetermined program of electrical stimulation to deep brain structures via implanted electrodes connected to a pulse generator\cite{bg3}. One major challenge with this procedure is to determine an optimal trajectory by placing the implant at the proper stimulation target while avoiding sensitive structures such as brain vessels \cite{can}. For epilepsy patients, ANT (the Anterior Nucleus of the Thalamus) is the stimulation target as it can influence the brain's predisposition to epileptic seizures\cite{bg4}. To standardize and facilitate placement and improve outcomes, it is desirable to develop an automatic algorithm for accurate ANT target localization.

The standard approach for automatic target localization is the atlas-based registration. However, this approach suffers from a lack of robustness when the anatomic differences between source and target images are large or when the number of atlases is limited \cite{atlas}. This can be of importance for our application due to the well documented variability in the ANT anatomy and thalamic atrophy caused by persistent epileptic seizures\cite{bg5}. Besides, classic registration methods are not very efficient as they require long processing time. Recently, a DL-based approach has been proposed to localize the ANT target via heatmap matching (HM)\cite{han}. In this method, heatmaps are generated by rendering a spherical 3D Gaussian centered on the coordinates of pseudo labels and the Mean Squared Error (MSE) loss between the output and synthetic heatmap is minimized during training. During inference, the final ANT targets are obtained from the model’s output by computing the argmax of the voxel probabilities. However, as suggested by Nibali \textit{et al.}\cite{dsnt}, the major limitation of HM is that gradient flow begins at the heatmap rather than the numerical coordinates, leading to a mismatch between the optimization objective, i.e., the similarity between heatmaps, and the evaluation metric for target localization, i.e., the distance between the coordinates of the predicted target and the ground truth. To overcome this limitation, a DSNT\cite{dsnt} (differentiable spatial to numerical transform)  layer has been proposed to fully preserve the spatial generalization as well as the end-to-end differentiability of the model. In this work, we conduct a comparative study to investigate the ANT target localization performances for DBS with different localization approaches including (1) traditional registration method (2) HM and (3) DSNT. For learning-based methods, we also explore the impact of the amount of training data on the localization performance. Lastly, we compare these results to the manual inter/intra-rater variability observed in a study conducted across two institutions.

\section{METHODS}
\subsection{Data and Materials}
Our dataset consists of 309 T1-weighted MRI images from a database of patients who underwent a DBS implantation for movement disorders, i.e., Parkinson Disease or Essential Tremor at Vanderbilt University. The in-plane resolution varies from 0.4356 to 1 mm and slice thickness from 1 to 6 mm. We randomly split the entire dataset into 259 images for training and validation and the rest 50 images for testing. The ANT target on one atlas from a different set is manually annotated by an experienced neurosurgeon. Atlas-based registration\cite{ants} is applied between the atlas image and each training sample to generate \textbf{pseudo labels}, i.e., 3D coordinates. For all images in our testing set, two experienced neurosurgeons from two different institutions are asked to annotate the ANT targets on both left and right thalamus. One of the neurosurgeons is asked to annotate the testing set twice for intra-rater variability analysis. The 3D coordinates of the annotated targets are collected using an open-source software ITK-SNAP \cite{itksnap}. For image preprocessing, we resample all the images to isotropic voxel size of $1\times1\times1$ mm$^{3}$ by trilinear interpolation and normalize the image intensities to the range of 0 and 1.

\subsection{Overview}
In this study, we explore several target localization methods for the ANT-DBS application including (1) traditional registration, (2) HM and (3) DSNT. We follow the two-stage strategy proposed in Liu \textit{et al.,}\cite{han} to localize the targets in a coarse-to-fine manner. The entire pipeline of our localization process is illustrated in Figure 2. Specifically, given an unseen testing image, or query image, we detect and crop a region of interest (ROI) around the thalamus in the first stage and localize the ANT target within the cropped ROI in the second stage. For the first stage, to reduce the inference time, we train a 3D U-Net\cite{unet} denoted as $F_{seg}$ to segment the left and right thalamus from downsampled MRI scans and then crop an ROI with a size of $64\times64\times64$ centered at the segmented thalamus. Once we collect the left and right ROIs of the entire dataset, we flip the left ROIs to the right such that all the cropped ROIs have consistent image contents. The consistency of input data reduces difficulty of the downstream task and is thus beneficial for our training. These ROIs are ready to be used as the input volumes for the second-stage localization network. In the following sections, we will discuss each localization method in detail.

\begin{figure}[t]
\includegraphics[width=1\columnwidth]{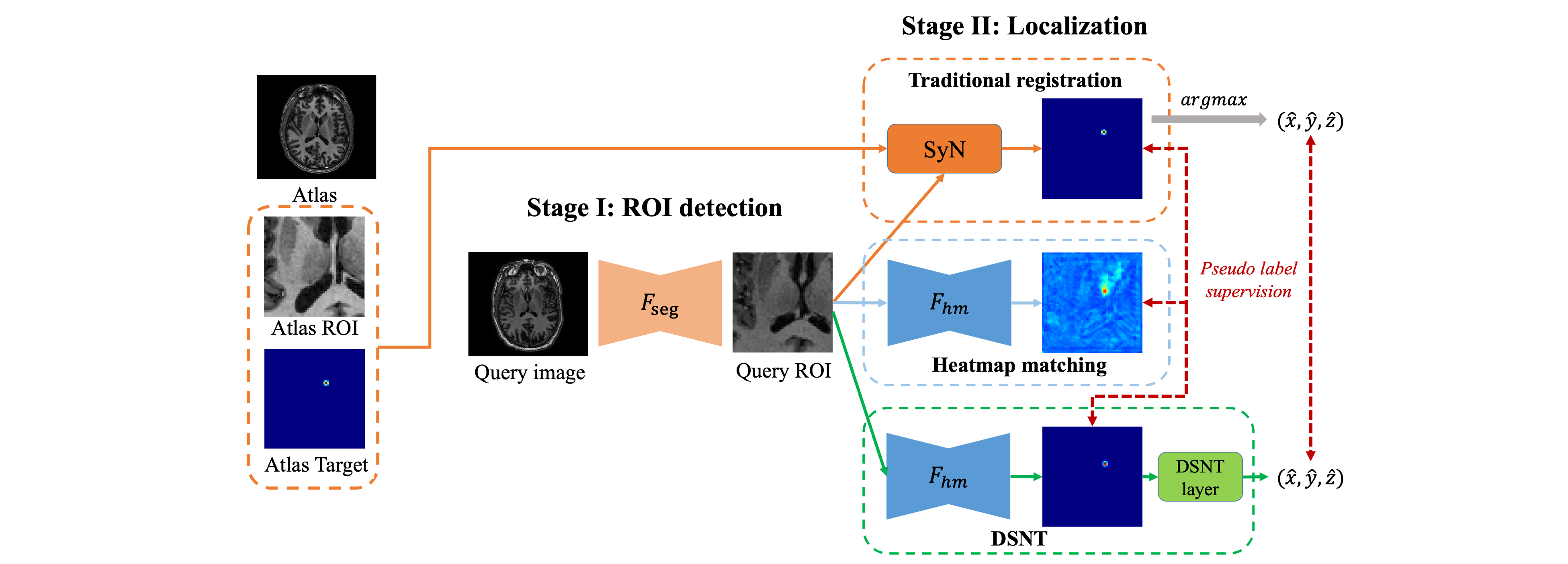}
\centering
\caption{An illustration of the two-stage target localization process and the localization methods involved in our comparative study. Note that the $F_{hm}$ in heatmap matching and DSNT share the same network architecture but do not share the weights.}
\label{fig2}
\end{figure} 

\subsection{Registration}
Image registration has been widely used to establish a spatial correspondence between biological structures across different images \cite{jianing}. Registration-based methods are useful for medical image analysis tasks when there are strong anatomical priors in the associated tasks. In our application, the ANT-DBS targets can be localized by registering our atlas ROI to the query ROI. Once the query ROI and the atlas ROI are well-aligned, the obtained deformation field can be used to project the ANT target on the atlas ROI, i.e., the one that is manually annotated by the expert, to the query ROI. Specifically, we use a well-accepted registration algorithm for brain MRI known as SyN\cite{syn} to register atlas ROI to each query ROI in the testing set. In implementation, we use the SyN with mutual information as optimization metric in the ANTs\cite{ants} package. 

\subsection{Heatmap matching}
For heatmap matching, we first generate synthetic heatmaps $y$ by rendering a spherical 3D Gaussian centered on the pseudo labels; their standard deviation is 1.5 mm. The maximum value of the heatmap is scaled to 1 and any value below 0.05 is set to 0. We train a 3D U-Net $F_{hm}$ to perform a per-voxel regression to perform heatmap matching. A query ROI $x$ is mapped to a probability map $F_{hm}(x)$, where $F_{hm}$ has the same network architecture as the one proposed in Liu \textit{et al.}\cite{han}. During training, a weighted MSE (WMSE) loss is used to minimize the heatmap difference between $F_{hm}(x)$ and $y$. Specifically, in the output probability map, we set the weights associated with positive entries and zero entries from pseudo label maps as $W_{p}$ and $W_{z}$, as described in Equation 1 and 2.

\begin{equation}
    W_{p}=\frac{N_{z}}{N_{p}+N_{z}}
\end{equation} 

\begin{equation}
    W_{z}=\frac{N_{p}}{N_{p}+N_{z}}
\end{equation}

where $N_{p}$ and $N_{z}$ are the number of positive and zero voxels in pseudo label map. During inference, the final predicted target $\widetilde{p}$ is obtained by taking the voxel with the highest activation value from the output probability map, as shown in Equation 3.
\begin{equation}
\widetilde{p}=\arg\max(F_{hm}(x))
\end{equation}

\subsection{DSNT}
To enable end-to-end differentiability for coordinate regression, we build a DSNT-based model $F_{dsnt}$ by adding a DSNT layer on top of the last layer of $F_{hm}$. The input of the DSNT layer is a normalized localization probability heatmap $F_{hm}(x)$ such that the values of the heatmap are non-negative and sum to 1. In our implementation, we use the softmax activation function to normalize the heatmap. Besides, we extend the original 2D implementation\footnote{\url{https://github.com/anibali/dsntnn}} of DSNT layer to 3D. Specifically, we create three matrices $X$, $Y$ and $Z$ which contain their own x, y, or z-coordinates and are scaled such that the top-left-front and bottom-right-back corner of the volume are at $(-1, -1, -1)$  at $(1, 1, 1)$, respectively. We can obtain the 3D coordinates $\widetilde{p}$ from the normalized heatmap as described in Equation 4.

\begin{equation}
    \widetilde{p}=[\langle{F_{hm}(x), X}\rangle_{F}, \langle{F_{hm}(x), Y}\rangle_{F}, \langle{F_{hm}(x), Z}\rangle_{F}]
\end{equation}

where $\langle{\cdot, \cdot}\rangle_{F}$ denotes the Frobenius inner product. During training, we minimize the MSE loss between the predicted coordinates $\widetilde{p}$ and the ground truth coordinates $p$ as well as a Jensen-Shannon divergence loss to minimize the distribution divergence between the generated heatmap and the ground truth Gaussian distribution, as described in Equation 5.

\begin{equation}
L_{DSNT}=L_{MSE}(\widetilde{p},p)+\alpha L_{JS}(F_{hm}(x),y)
\end{equation}

where $\alpha$ is set to be 1 empirically. During inference, we take the argmax of the generated heatmap as our final target because we find that the argmax of the heatmap is slightly more reliable than the coordinate output on our validation set.

\subsection{Implementation Details}
For pre-processing, we normalize the intensity of each ROI to the range of 0 and 1 before feeding them to the networks. We apply data augmentation to avoid overfitting and improve model generalization ability. Specifically, we augment our training data by random intensity shifts in the range of 0.8 and 1.2, random scaling in the range of 0.9 to 1.1 and random rotation in the range of $-10$ to $10$ degrees along each axis. The best hyperparameters are determined by grid-search within the range of $10^{-2}$ to $10^{-5}$. The best hyperparameters are selected based on the localization performance on the validation set. For training, we use the Adam optimizer \cite{adam} with a weight decay of $10^{-4}$ and a batch size of 16. The CNNs are implemented in PyTorch \cite{pytorch} and MONAI on a Ubuntu desktop with an NVIDIA RTX 2080 Ti GPU. During inference, we first resample the ROI of the testing image to isotropic voxel size of $1\times1\times1$ mm$^{3}$ and obtain the prediction on this resampled image. The actual coordinates of the target are computed based on the resampling factor.

\section{EXPERIMENTS and RESULTS}
\subsection{Experimental Design}
In our experiments, we compare the localization results obtained by each method against the ground truth annotated by the more experienced neurosurgeon. The labels annotated by the same neurosurgeon but at a different time and the other neurosurgeon are used to compute the intra- and inter-rater variability respectively, which can be viewed as the lower bounds of the localization error for the automatic localization algorithms. For quantitative evaluation, we use the mean radial error (MRE) to measure the Euclidean distance between prediction and ground truth as well as the successful detection rate (SDR) for three radii (2 mm, 4 mm and 6 mm). We also compare the inference time per testing image and the impact of the amount of training data for each method. 

\subsection{Results}
In Table 1, we show the MRE and SDR achieved by different methods with 100\%, 50\% and 25\% of the training data to study the effect of the training set size on the results. The intra-rater and inter-rater variability of MRE are $2.04\pm0.87$ mm and $2.42\pm1.17$ mm, respectively. Our results show that when trained with 100\% of the training data, both the DL-based methods, i.e., HM and DSNT, and the registration method SyN have achieved MRE comparable to the inter/intra-rater variability. We perform the paired t-test with a p-value of 0.05 and find no statistically significant difference between each method. In terms of SDR, HM leads to accurate target localization, i.e., targets within 2 mm, more often than SyN and DSNT. We also observe that the SDR achieved by HM and registration are comparable to the inter-rater performance. On the other hand, when trained with less training data, e.g., with only 25\% of training data, we observe that DSNT can yield better localization performance than HM as it leads to significantly lower mean MRE with lower standard deviation. Lastly, in terms of inference time, we show that the DL-based methods can reduce the inference time by orders of magnitude compared to the registration method.

\begin{table}[h]
\centering
\caption{Comparison of localization performances achieved by different methods}
\begin{tabular}{|c|ccc|ccc|c|}
\hline
\multirow{2}{*}{METHODS} & \multicolumn{3}{c|}{MRE($\downarrow$) (mm)}           & \multicolumn{3}{c|}{SDR($\uparrow$) (\%)}             & \multirow{2}{*}{Time (sec)} \\ \cline{2-7}
                         & 100\%     & 50\%      & 25\%       & 100\%        & 50\%      & 25\%      &                             \\ \hline
Intra-rater              & 2.04±0.87 & -         & -          & 58, 99, 100  & -         & -         & -                           \\
Inter-rater              & 2.42±1.17 & -         & -          & 45, 93, 100  & -         & -         & -                           \\
SyN                      & 2.49±1.09 & -         & -          & 40, 91, 100  & -         & -         & 3.7                         \\
HM                       & 2.34±1.12 & 4.38±3.56 & 7.13±10.17 & 42,  92, 100 & 5, 53, 91 & 1, 43, 85 & 0.005                       \\
DSNT                     & 2.45±1.03 & 4.23±1.98 & 4.17±1.46  & 29, 95, 100  & 3, 50, 95 & 2, 47, 87 & 0.005                       \\ \hline
\end{tabular}
\end{table}

\section{BREAKTHROUGH WORK}
We have compared several localization methods for ANT-DBS targeting and assessed the performances against the annotations from multi-raters. Our experimental result show that the DL-based localization models can lead to an accuracy that is comparable to the measured intra/inter-rater variability. These models can be trained with pseudo labels that are generated by only one expert-annotated label and can make inference much faster than traditional registration methods. To the best of our knowledge, this work is the first to investigate various localization algorithms for ANT-DBS with multi-rater annotations.

\section{DISCUSSION and CONCLUSION}
An interesting observation from our experiment is that the DSNT-layer can improve localization performance when the model is trained with a small amount of data. We conjecture that the outperformance of DSNT might be due to the MSE loss term computed on the 3D coordinates. In our second stage, the ROIs are adjusted to have consistent image contents and thus the 3D coordinates of pseudo label targets do not vary much. With the additional supervision from the coordinate loss, the model equipped with a DSNT layer is more likely to converge faster than the HM model when the amount of training data is limited. With a larger amount of training data, HM is shown to be a better localization algorithm than DSNT as it produces more high-quality targets. In summary, in this work, we perform a multi-rater comparative study for ANT-DBS target localization. Our experimental results demonstrate that DL-based models that are trained with \textbf{pseudo labels} can achieve a localization accuracy that is comparable to the intra/inter-rater variability of experienced neurosurgeons with much faster inference time than traditional method. A validation study in which surgeons are asked to assess the quality of the localization, i.e., whether it could be used clinically is ongoing.

\section{Acknowledgments}
This work has been supported by NIH grant R01NS095291 and by the Advanced Computing Center for Research and Education (ACCRE) of Vanderbilt University. The content is solely the responsibility of the authors and does not necessarily represent the official views of these institutes.

\end{document}